# Active control of broadband plasmon-induced transparency in terahertz hybrid metal-graphene metamaterial

Zhaojian Zhang,[a] Junbo Yang,*[b] Xin He,[b] Yunxin Han,[b] Jingjing Zhang,[a] Jie Huang,[a] Dingbo Chen[a] and Siyu Xu[a]

[a] College of Liberal Arts and Sciences, National University of Defense Technology, Changsha 410073, China. E-mail: 376824388@qq.com

[b] Center of Material Science, National University of Defense Technology, Changsha 410073, China. E-mail: yangjunbo@nudt.edu.cn

**Abstract**

A hybrid metal-graphene metamaterial (MM) is reported to achieve the active control of the broadband plasmon-induced transparency (PIT) in THz region. The unit cell consists of one cut wire (CW), four U-shape resonators (USRs) and monolayer graphene sheets under the USRs. Via near-field coupling, broadband PIT can be produced through the interference between different modes. Based on different arrangements of graphene positions, not only can we achieve electrically switching the amplitude of broadband PIT, but also can realize modulating the bandwidth of the transparent window. Simultaneously, both the capability and region of slow light can be dynamically tunable. This work provides schemes to manipulate PIT with more degrees of freedom, which will find significant applications in multifunctional THz modulation.

## 1. Introduction

Electromagnetically-induced transparency (EIT) is a quantum phenomenon owing to the interference between two different excitation pathways in a three-level atomic system,[1] and the strong dispersion within the transparent window can promise for the applications in slow light and enhancing nonlinear optical effect.[2,3] However, the experimental conditions of EIT are harsh such as stable pumping and low temperature,[1] which restrict its further extensive application. Besides, the EIT effect is mainly observed in optical frequency range due to the limited energy interval for transitions of quantum states, which makes it hard to produce EIT at terahertz (THz) region.[4] Recently, as an EIT analogue in a plasmonic system, plasmon-induced transparency (PIT) has been proposed and widely realized in THz band based on metamaterials (MMs).[5] The metal meta-resonators of such MMs, including split-ring resonators (SRRs),[6,7] U-shape resonators (USRs) [8,9] and cut wires (CWs),[10,11] can support electromagnetic resonances at THz region. Via near-field coupling of resonant elements, PIT can be realized from the destructive interference between the bright mode (directly excited mode) and dark mode (indirectly excited mode) [12] or two bright mode.[13] Nowadays, MMs-based PIT has been found a wide range of applications as THz filters,[14] slow light devices [15] and sensors.[16]

Although PIT can be tuned by changing geometric parameters of MMs, it's essential to propose a dynamic tunable PIT in practice. Superconductor is used to obtain temperature modulation of transparent window position,[17,18] and utilizing photoconductive silicon island, an on-to-off switch of PIT also can be realized.[19] However, these modulation schemes suffer from relative slow response time (approximately on the order microsecond and millisecond, respectively). As a two-dimension (2D) material, graphene has drawn more attention owing to the superior photoelectric modulation performance. Due to the high mobility of graphene electrons, the Fermi energy of graphene can be tuned via electrical gating or chemical doping, bringing about the flexible change of the graphene surface conductivity according to Kubo formula.[20] The corresponding response time is on the order of picosecond, promising for the ultrafast switching.[21] So far, plenty of works have concentrated on proposing PIT with tunable transparent position based on graphene nanopattern MMs [22,23] or graphene complementary MMs [24,25] in THz regime. Nevertheless, extra noises will be involved at adjacent frequency spectra during such manipulation process. Recently, the hybrid metal-graphene MMs are introduced to achieve active control of the transparency window amplitude.[26,27] However, these works only focus on switching the amplitude of a narrowband PIT. The modulation with more degrees of freedom need to be demonstrated.

In this paper, we report a hybrid metal-graphene MM to achieve the active control of the broadband PIT in THz region. The unit cell consists of one CW, four USRs and monolayer graphene sheets. Via near-field coupling, PIT can be produced through the destructive interference of modes, the broadband of which can be theoretically explained by the plasmonic hybridization model (PHM). Most importantly, based on different graphene positions, not only can we achieve on/off switching of broadband PIT, but also can manipulate the bandwidth of the transparent window by electrically tuning the Fermi energy of graphene sheet. Such PIT with tunable bandwidth is introduced for the first so far as we know. Simultaneously, both the capability and region of slow light can be dynamic tunable. This work makes possible the PIT modulation with more degrees of freedom, which will find significant applications in multifunctional modulators, switches and buffers for future THz wireless interconnects.

## 2. Structures and materials

The top (from z direction) and front (from y direction) view for the unit cell of the proposed hybrid metal-graphene MM are shown in Fig. 1(a) and (b), the corresponding detailed geometric parameters are given in the caption. As shown in Fig. 1, a pair of USRs are arranged on each side of one CW, and two sheets of monolayer graphene are laid under the USR pairs, respectively. The substrate is chosen as quartz at the bottom. There is a periodic structure in the x and y direction.

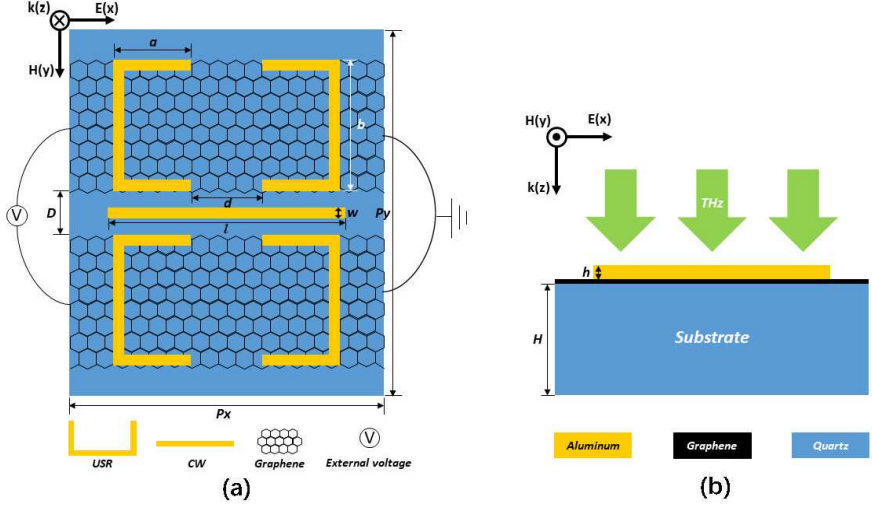

Fig.1 (a) The top view (from z direction) for the unit cell of the MM. The geometrical parameters: $a$ = 28 μm, $b$ = 48 μm, $d$ = 26 μm, $w$ = 4 μm, $l$ = 86 μm, $D$ = 10 μm, $P_x$ = 114 μm, $P_y$ = 134 μm. (b) The front view (from y direction) for the unit cell of the MM. The geometrical parameters: $h$ = 200 nm, $H$ = 40 μm.

The USRs and CW are both made of aluminum, of which conductivity is set as $\sigma_{Al} = 3.72 \times 10^7\ S \cdot m^{-1}$.[28] The quartz substrate is described as a lossless dielectric medium with relative permittivity $\varepsilon_{sub} = 3.76$. The monolayer graphene is modeled as an 2D conductive sheet, the surface conductivity of which is given by $\sigma_{gra} = \sigma_{inter} + \sigma_{intra}$, where $\sigma_{inter}$ and $\sigma_{intra}$ represent interband and intraband conductivity of graphene, respectively. At the lower terahertz region, the interband contributions are forbidden by Pauli blocking principle,[26] therefore, the graphene conductivity can be derived by Kubo formula:[29]

$$\sigma_{gra} \approx \sigma_{intra}(\omega, E_F, \Gamma, T) = i \frac{e^2 k_B T}{\pi \hbar^2 (\omega + i\Gamma)} [\frac{E_F}{k_B T} + 2 ln(e^{-\frac{E_F}{k_B T}} + 1)] \qquad (1)$$

Where $e$ is the unit charge, $\hbar$ is the reduced Planck's constant, $T$ is environment temperature in the unit of K, $k_B$ represents the Boltzmann constant, and $E_F$ is the Fermi energy. $\Gamma = 1/\tau$ is the carrier scattering rate, while $\tau$ is the carrier relaxation time which is described as $\tau = \mu E_F / e v_F^2$ with $\mu$ being the carrier mobility and $v_F$ being the Fermi velocity. Here, we assume that $\mu = 3000\ cm^2/V \cdot s$ and $v_F = 1.1 \times 10^6\ m/s$, which are consistent with the experimental measurements.[30,31]

The Fermi energy of graphene can be tuned by external voltage, which can be realized based on gate electrodes.[32] In this structure, electrodes can be set on both sides of the MM to connect the graphene strips as shown in Fig. 1(a), which is similar to that scheme.[26] The manufacturing process of such MM can be described as follows:[33] First, the monolayer graphene can grow on copper substrates by chemical vapor deposition (CVD) and will be transferred onto the quartz substrate by dry transfer technique. Next, the graphene sheet is fabricated into strips by electron beam lithography (EBL) and oxygen plasma etching (OPE). Then, the aluminum layer is deposited on the top of the existing structure. Finally, utilizing EBL and OPE, both the metal resonant elements and metal electrode pairs can be manufactured.

The finite-difference time-domain (FDTD) solution is utilized for simulating features of introduced device with periodical boundary conditions in the x and y directions and perfectly matched layer (PML) in the z direction, moderate mesh grid is employed to ensure the accuracy. A THz source with x-direction polarization is put on the top of the MM and propagates along the z direction, a monitor is set at the bottom of substrate to detect the transmitted power.

## 3. Results and discussions

At first, three kinds of MMs without graphene are simulated to investigate the mechanism of PIT, the unit cell of each MM only consists of a CW, a USR and a CW coupled with one USR, respectively. The corresponding transmission spectrum and electric field distributions are shown in Fig. 2(a) and (c-e). When there is only a CW under the source with x-direction polarization, a symmetric Lorenz-type resonance appears around 1 THz as exhibited in Fig. 2(a), which is a typical dipolar localized surface plasmons (DLSPs) as shown in Fig. 2(c).[34] The inductive–capacitive (LC) resonance can be supported by the USR [35] but cannot be efficiently excited by such polarized light as depicted in Fig. 2(a) and (d). However, when combining the USR with one CW, the LC resonance will be motivated by DLSPs as shown in Fig. 2(e). Here, DLSPs are directly excited by the source, acting as the bright mode. LC resonance is indirectly excited by DLSPs, which is the dark mode. Through the destructive interference between the bright and dark mode, PIT will come out near the common resonant frequency of two modes as shown in Fig. 2(a).

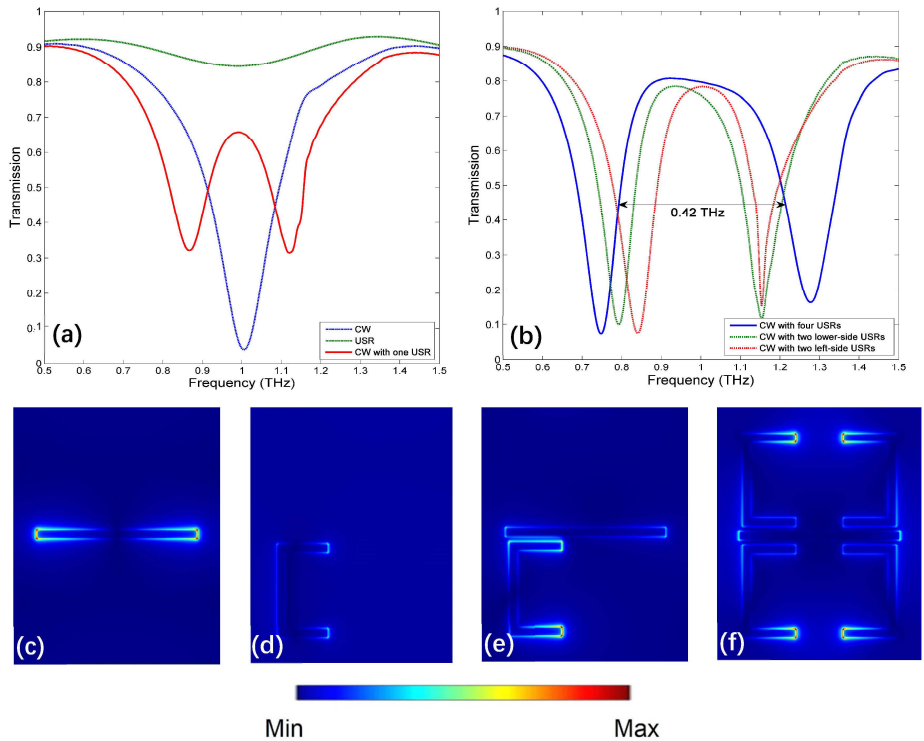

Fig. 2 (a) The transmission spectrum corresponding to the unit cell composed of a CW, a USR and a CW coupled with one USR, respectively. (b) The transmission spectrum corresponding to the unit cell composed of a CW coupled with four USRs, a CW coupled with two lower-side USRs and a CW coupled with two left-side USRs, respectively. (c) The electric field distribution at 1 THz when only one CW. (d) The electric field distribution at 1 THz when only one USR. (e) The electric field distribution at 1 THz when one CW is coupled with one USR. (f) The electric field distribution at 1 THz when one CW is coupled with four USRs.

The transmission spectrum of MM composed of one CW coupled with four USRs (the structure in Fig. 1 without graphene) are shown in Fig. 2(b), and the full wave at half maximum (FWHM) of the transparent window is broadened to 0.42 THz, the corresponding electric filed distribution is presented in Fig. 2(f), which shows that LC resonances are simultaneously supported by four USRs. To clarify the mechanism of broadband transparent window, three sets of unit cells are investigated: a CW coupled with one USR, a CW coupled with two USRs on the lower side and a CW coupled with two USRs on the left side, the corresponding z-component distributions of electric field within the transparent window are shown in Fig. 3(a), (b-c) and (d-e), respectively. When there is single coupled USR, the transparent window is narrow, of which FWHM is 0.16 THz as shown in Fig. 2(a). Fig. 3(a) indicates that only one kind of dark mode exists on the USR at 1 THz. When two USRs are arranged on the lower side, the FWHM is extended to 0.28 THz as depicted in Fig. 2(b). According to Fig. 3(b-c), the hybridization of double LC resonances on the USR pair will produce two dark supermodes with opposite electric field directions via electromagnetic interaction, which can be explained by the PHM.[36] The two supermodes are extremely close in resonant frequencies, consequently forming a relatively broadband PIT. Same procedure can be applied when USR pair are on the left side as shown in Fig. 3(d-e). Finally, when it comes to four USRs, the dark supermodes in Fig. 3(f-g) can be regard as the hybridization of four kinds of dark supermodes supported by USR pairs as given in Fig. 3(b-e), therefore, producing a wider transparent window with the FWHM of 0.42 THz.

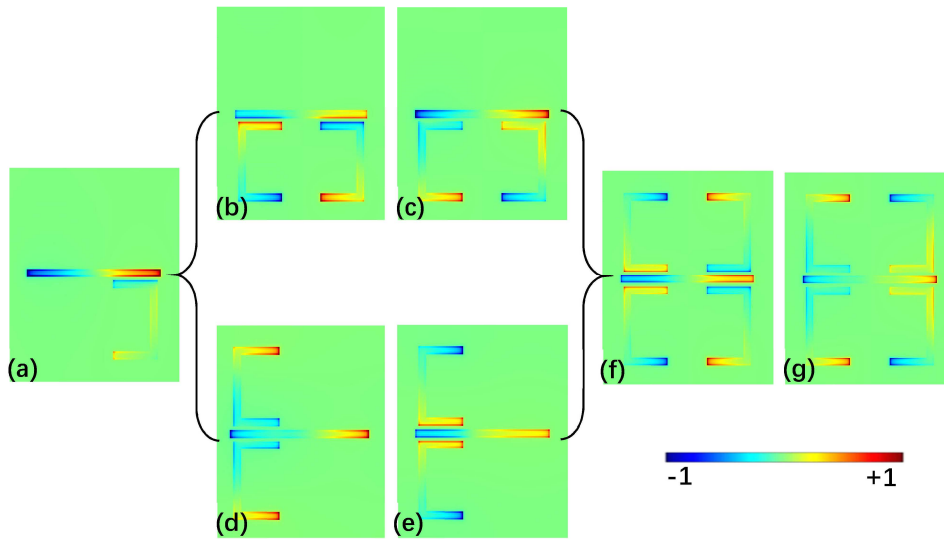

Fig. 3 The mode ($E_z$) distribution corresponding to (a) A CW coupled with one USR at 1 THz. (b-c) A CW coupled with two USRs on the lower side at 0.9 THz and 1 THz, respectively. (d-e) A CW coupled with two USRs on the left side at 1.05 THz and 0.95 THz, respectively. (f-g) A CW coupled with four USRs at 0.95 THz and 1.05 THz, respectively.

Next, two monolayer graphene strips are integrated under the USR pairs respectively as shown in Fig. 1(a). The transmission spectrum corresponding to Fermi energy from 0.05 eV to 0.45 eV are shown in Fig. 4(a), such Fermi energy can be easily reached by applying external voltage.[37] As the Fermi energy increases, the amplitude of transparent window will reduce without central transparent frequency shift, consequently the adjacent frequency spectra will not be influenced. To assess the manipulation performance, the modulation depth (MD) is introduced:

$$MD = \frac{\Delta T}{T_0} \times 100\% \tag{2}$$

Where $T_0$ is the transmission of the transparent window center when Fermi energy is at 0.05 eV, $\Delta T$ is the central transmission change compared with $T_0$, and the corresponding details are collected in Table. 1. When Fermi energy is at 0.05 eV, the broadband transparent window still exists with central transmission 75%. However, the central transmission will decrease with Fermi energy rising. When it comes to 0.35 eV, the PIT effect almost vanishes. Then the transmission spectrum will evolve into a symmetric Lorenz profile at 0.45 eV with the central valley value 21%, the corresponding MD can reach 72%, indicating that such MM has an excellent performance for on/off electrical switching of broadband PIT. Simultaneously, this device possesses the ultrafast response time owing to the characteristics of graphene.

As one of the most important applications of PIT, slow light is also investigated under different Fermi energy. The slow light can be qualified by the delay time $\tau_g$:[26]

$$\tau_g = \frac{d\psi(\omega)}{d\omega} \tag{3}$$

Where $\psi(\omega)$ stands for the transmission phase shift from the light source to the monitor. As shown in Fig. 4(b), when Fermi energy is at 0.05 eV, there is a slow light region duo to the extreme dispersion within the transparent window. As Fermi energy increases, the PIT effect will be weaker, consequently losing the slow light capability. The details are shown in Table. 1. Therefore, the electrical control of broadband slow light ban be realized.

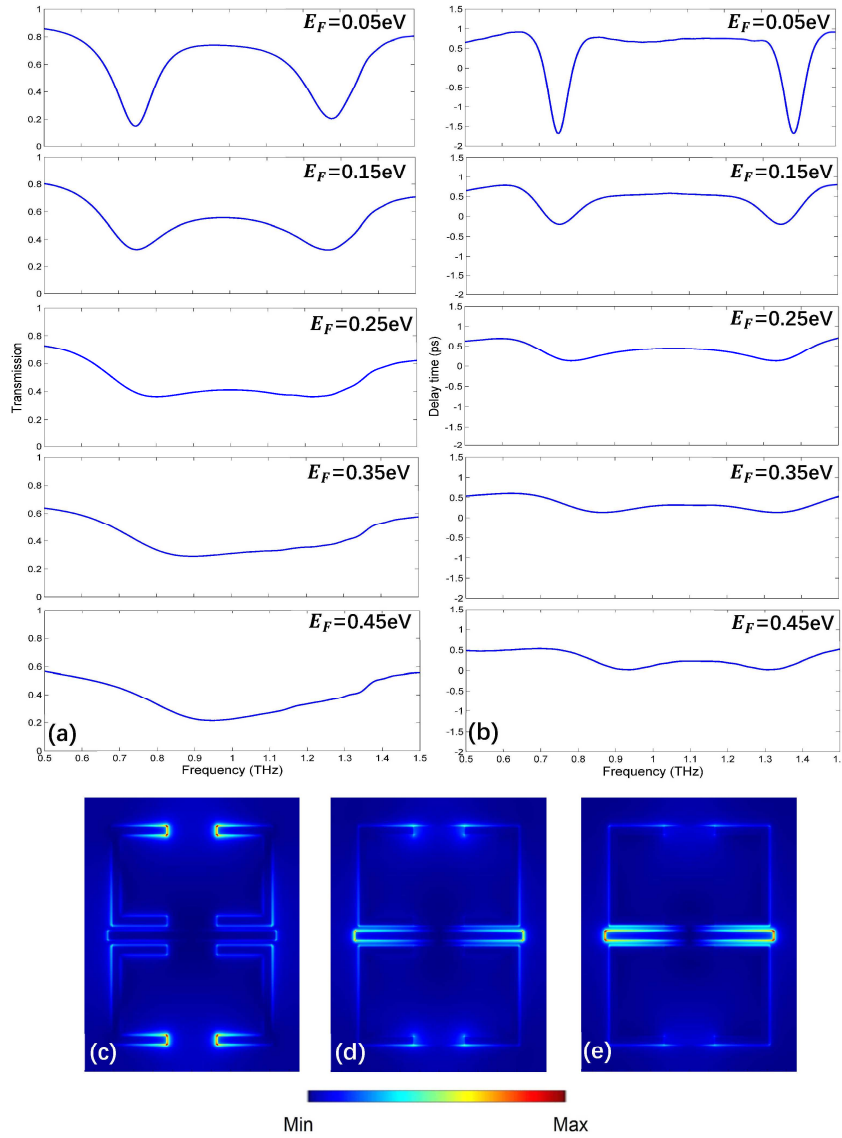

Fig. 4 (a) The transmission spectrum corresponding to Fermi energy from 0.05 eV to 0.45 eV. (b) The slow light region corresponding to Fermi energy from 0.05 eV to 0.45 eV. (c-e) The electric filed distribution of MM at 0.05 eV, 0.25 eV and 0.45 eV, respectively.

Table. 1 The details of the PIT transmission, MD and delay time at different Fermi energy.

| Fermi energy | 0.05 eV | 0.15 eV | 0.25 eV | 0.35 eV | 0.45 eV |
|---|---|---|---|---|---|
| PIT transmission | 0.75 | 0.56 | 0.41 | 0.29 | 0.21 |
| MD | 0 | 25.3% | 45.3% | 61.3% | 72% |
| Delay time | 0.76 ps | 0.58 ps | 0.44 ps | 0.31 ps | 0.22 ps |

The electric field distributions of the unit cell at different Fermi energy are shown in Fig. 4(c-e). It is obviously that the dark modes supported by USR pairs will be suppressed as Fermi energy increases. Because at THz region, graphene will tend to be more metallic with the rising of Fermi energy, consequently bringing about more energy loss on the surface.[38] As electric fields on USR pairs fade, the destructive interference between bright and dark mode will be cancelled, finally there will be only bright mode at 0.45 eV as shown in Fig. 4(e), the corresponding symmetric Lorenz transmission spectrum in Fig. 4(a) also prove this point.

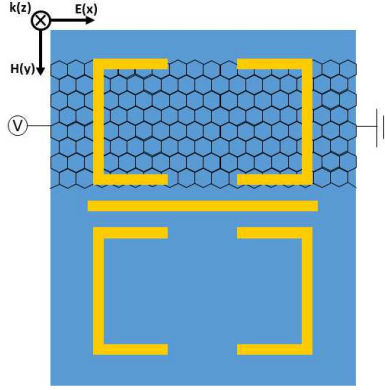

Fig. 5 The top view (from z direction) for the unit cell of the MM which can control the bandwidth of PIT.

Furthermore, via integrating monolayer graphene under one of USR pairs as depicted in Fig. 5, the bandwidth of PIT also can be electrically tunable. As shown in Fig. 6(a), when Fermi energy increases, the bandwidth of PIT will be narrower. To qualify such manipulation, the modulation width (*MW*) is utilized:

$$MW = \frac{\Delta F}{F_0} \times 100\% \qquad (4)$$

Where $F_0$ is the FWHM of the transparent window when Fermi energy is at 0.05 eV, $\Delta F$ is the change of FWHM relative to $F_0$, the details are shown in Table. 2. When starting at 0.05 eV, The PIT has a broadband transparent window with the FWHM of 0.41 THz. However, the FWHM will become narrower with the rising of Fermi energy. From 0.05 eV to 1.25 eV, PIT has an evolution from broadband to narrowband, the *MW* is up to 63.4%. Meanwhile, the central transparent frequency still remains at 1 THz, which has no shift compared with the original broadband PIT. Moreover, the region of slow light is also tuned from broadband to narrowband as shown in Fig. 6(b). As Fermi energy increases, a narrower PIT will lead to a narrower slow light, and the delay time can be lifted from 0.78 ps to 1.56 ps at 1 THz as given in Table. 2. Such PIT with dynamic tunable bandwidth can achieve controlling PIT with more degrees of freedom, and, to the best of our knowledge, is proposed for the first time. It has to mention that such Fermi energy level can also be fulfilled by gated voltage.[37]

The corresponding electric field distributions at different Fermi energy are shown in Fig. 6(c-e). As Fermi energy rises, the USR pair upper the graphene will suffer from the energy loss caused by graphene surface, therefore the corresponding supported dark mode will be suppressed. According to PHM, the lack of dark modes will bring about the narrowing of bandwidth. A narrower and sharper profile of PIT indicates a more intense dispersion within the transparent window,[39] consequently the slow light capability is enhanced. However, the PIT peak value and bandwidth are smaller compared with the spectrum given in Fig .2(b), which can attribute to the loss caused by the metal-like graphene. What's more, if we put graphene strips under both USR pairs and apply independent voltage on each graphene,[40] not only the narrowband PIT can be switched from on to off state, but also the amplitude and bandwidth of the transparent window can be controlled simultaneously at the same MM.

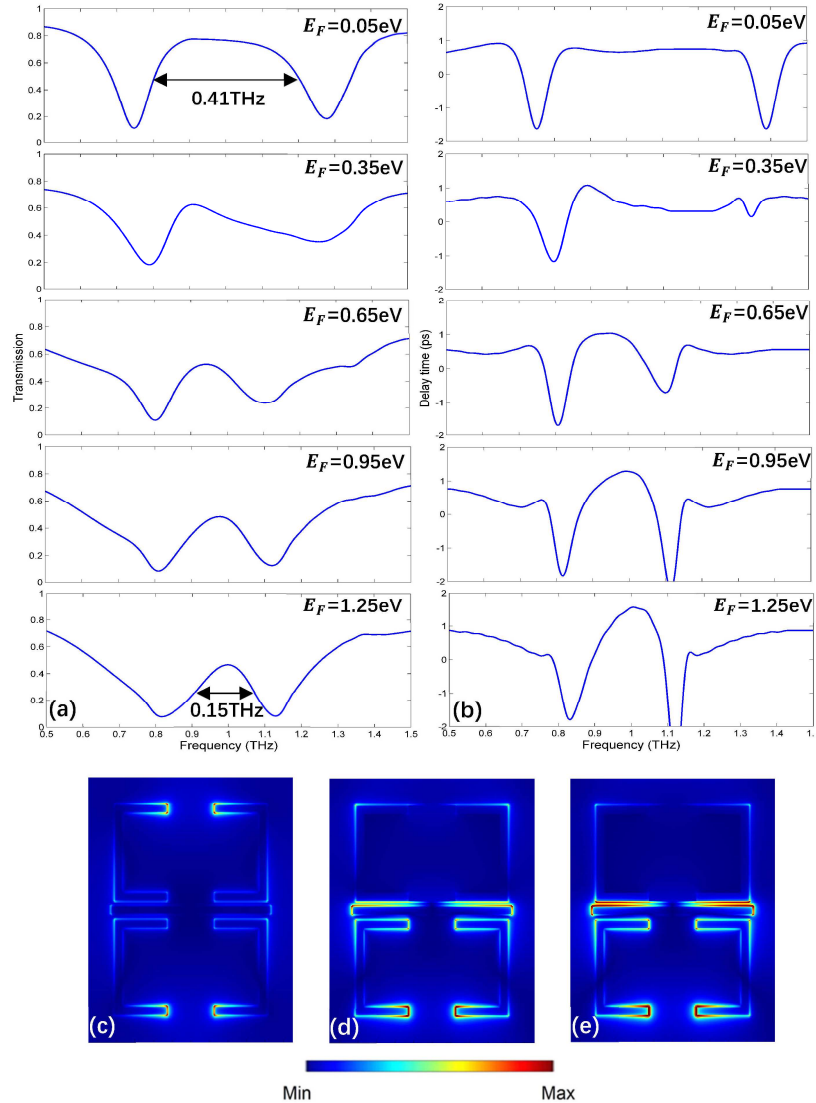

Fig. 6 (a) The transmission spectrum corresponding to Fermi energy from 0.05 eV to 1.25 eV. (b) The slow light region corresponding to Fermi energy from 0.05 eV to 1.25 eV. (c-e) The electric filed distribution of MM at 0.05 eV, 0.65 eV and 1.25 eV, respectively.

**Table. 2 The details of the PIT FWHM, MW and delay time at different Fermi energy.**

| *Fermi energy* | 0.05 eV | 0.35 eV | 0.65 eV | 0.95 eV | 1.25 eV |
|---|---|---|---|---|---|
| *PIT FWHM* | 0.41 THz | 0.33 THz | 0.19 THz | 0.18 THz | 0.15 THz |
| *MW* | 0 | 19.5% | 53.7% | 56.1% | 63.4% |
| *Delay time* | 0.78 ps | 1.07 ps | 1.04 ps | 1.29 ps | 1.56 ps |

## 4. Conclusion

In summary, we propose a metal-hybrid MM to realize broadband PIT with electrical tunable amplitude and bandwidth of the transparent window. The function can be achieved under different arrangements of graphene positions or applying different independent voltages on each graphene strip. The simulation results show that such MM not only has a significant performance on depth and width modulation of PIT, but also can realize electrically controlling both the capability and frequency region of slow light. Obviously, this device will play an important role in THz filtering, sensing, buffering and switching for future THz interconnects.

**Conflicts of Interest:** The authors declare no conflict of interest.

**Acknowledgments:** This work is supported by the National Natural Science Foundation of China (60907003, 61671455); the Foundation of NUDT (JC13-02-13, ZK17-03-01), the Hunan Provincial Natural Science Foundation of China (13JJ3001), and the Program for New Century Excellent Talents in University (NCET-12-0142).